# Photonic RF Channelization Based on Microcombs

Weiwei Han, Zhihui Liu, Mengxi Tan, Chaoran Huang, Jiayang Wu, Kun Xu, David J. Moss, and
Xingyuan Xu

*Abstract*— In recent decades, microwave photonic channelization techniques have developed significantly. Characterized by low loss, high versatility, large instantaneous bandwidth, and immunity to electromagnetic interference, microwave photonic channelization addresses the requirements of modern radar and electronic warfare for receivers. Microresonator-based optical frequency combs are promising devices for photonic channelized receivers, enabling full advantage of multicarriers, large bandwidths, and accelerating the integration process of microwave photonic channelized receivers. In this paper, we review the research progress and trends in microwave photonic channelization, focusing on schemes that utilize integrated microcombs. We discuss the potential of microcomb-based RF channelization, as well as their challenges and limitations, and provide perspectives for their future development in the context of on-chip silicon-based photonics.

*Index Terms*— Optical microcombs, RF channelization, microwave photonic.

## I. INTRODUCTION

WIDEBAND microwave signal receivers are used in a wide variety of modern electronic warfare systems, deep space tracking and telecommunications applications. Channelized reception refers to dividing the spectrum of the target signal, especially in the radio frequency (RF), microwave and millimeter wave bands, into different subchannels to achieve synchronous reception [1], [2], [3], [4], [5], [6], [7]. In general, the receiver's bandwidth, sensitivity, spectral resolution, and real-time processing speed are more demanding for complex signals with multiple formats, wide spectrums, and overlapping time domains in the electromagnetic environment. Initially, in the early days, the RF signal was demodulated by a superheterodyne receiver, but was suffered from image frequency interference [8]. Subsequently, channelized receivers based on electrical filter banks were proposed to channelize the signal in the frequency domain. While this technique has its advantages, receiving high-frequency broadband signals often involves multi-stage frequency conversion, massive filters, mixers and local oscillators (LOs) that can introduce electromagnetic interference. Although high-frequency electronic devices have been developed, they are ultimately limited by analog-to-digital / digital-to-analog converters (ADCs/DACs), resulting in a receiver operation bandwidth of only a few GHz, which is insufficient for modern requirements. In contrast, microwave photonic channelized receivers map the wideband signal onto the optical domain, and then divide it into multiple narrowband subchannels for parallel reception and processing, thus breaking through traditional electronic bottleneck limitations to achieve operating bandwidths from 1 to 10's of gigahertz. Benefiting from a high dynamic range, low power consumption, reconfigurability, and immunity to electromagnetic interference, photonic-assisted channelization has experienced unprecedented growth over the past two decades.

Initially, channel division for microwave photonic channelized receivers was achieved using surface acoustic wave (SAW) phased array diffraction [9], fiber Bragg gratings, and Fabry-Perot (FP) etalons. To reduce the system volume, it was proposed to integrate two Bragg gratings with a FP etalon [10]. Alternatively, optical frequency combs (OFCs) have emerged as attractive multi-wavelength Los that can greatly increased the available number of channels. Generally, four-wave mixing (FWM) or time lens techniques [ref] in optical fiber can be exploited for spectral broadening to create an optical frequency comb [11], [12], after which the RF signal can be broadcast onto each comb line, immediately followed by channel segmentation by periodic optical filters. From this, dual coherent optical combs were demonstrated [13], where one serves to multicast RF signals and the other acts as a LO, after which both are fed into the in-phase/quadrature (I/Q) coherent demodulator simultaneously. Spectral slicing can be performed using the vernier effect arising from a slight difference in the frequency spacing of two combs, eliminating the need for optical filters. Clearly, the stability and accuracy of the filter are not specified, which reduces the complexity of the filter. In recent years, wavelength scanning has been achieved by

Manuscript received XXX X, XXXX; revised XXX X, XXXX; accepted XXX X, XXXX. Date of publication XXX X, XXXX; date of current version XXX X, XXXX.

Corresponding author: Kun Xu, Xingyuan Xu.

W. Han, Z. Liu, X. Xu, K. Xu are with the State Key Laboratory of Information of Photonics and Optical Communications, School of Electronic Engineering, Beijing University of Posts and Telecommunications, Beijing 100876, China (e-mail: hanww@bupt.edu.cn; zhihuiliu@bupt.edu.cn; xingyuanxu@bupt.edu.cn; xukun@bupt.edu.cn).

M. Tan is with School of Electronic and Information Engineering, Beihang University, Beijing 100191, China (e-mail: simtan@buaa.edu.cn).

C. Huang is with Department of Electronic Engineering, The Chinese University of Hong Kong, Hong Kong, China (email: crhuang@ee.cuhk.edu.hk)

J. Wu, D. J. Moss is with Optical Sciences Centre, Swinburne University of Technology, Hawthorn, VIC 3122, Australia (e-mail: dmoss@swin.edu.au).

Color versions of one or more of the figures in this letter are available online at http://ieeexplore.ieee.org.

Digital Object Identifier 

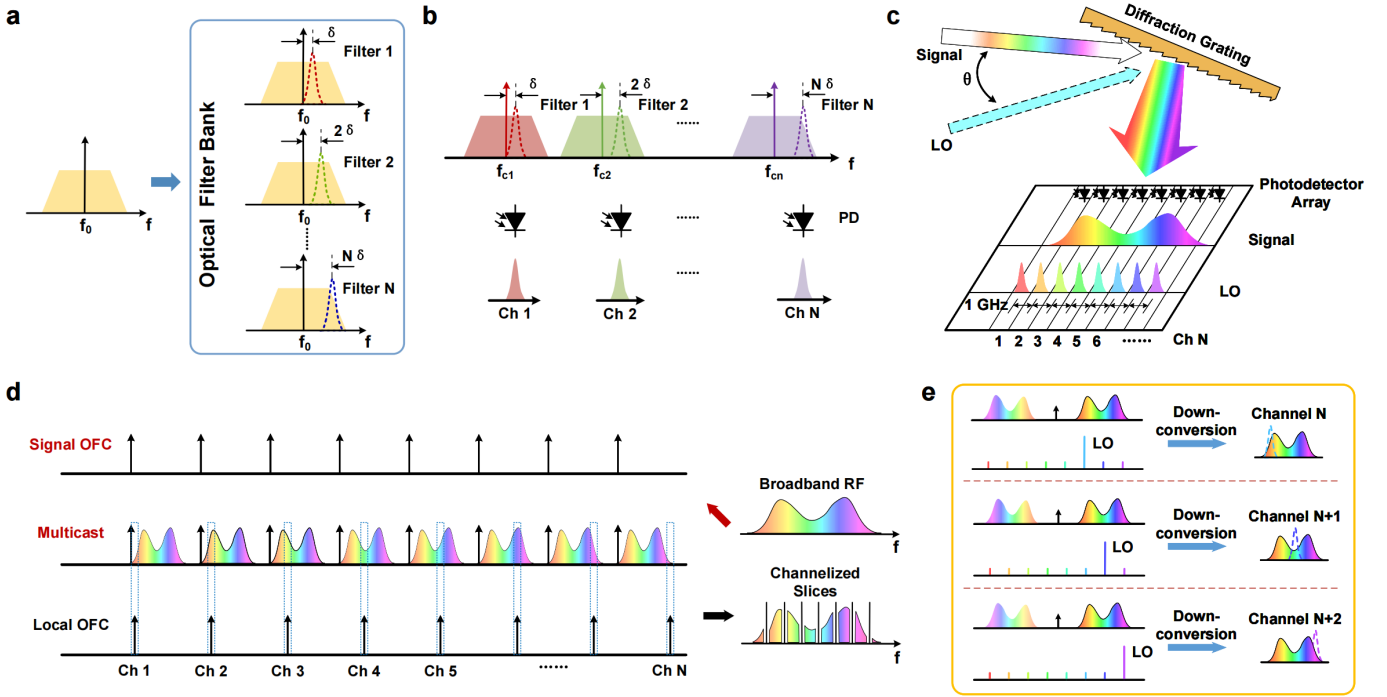

Fig. 1. Schematic diagrams of different microwave photonic channelized receiver schemes. (a) Optical filter bank. (b–c) Multi-wavelength laser sources. (d) Dual optical frequency combs. (e) Frequency shifting/scanning. Figures are adapted from [40], [41], [13], [69].

external laser source injection locking [14], acousto-optic frequency shifters (AOFS) [15], and optical switches [16]. Each wavelength marks a specific RF channel, which is then received serially by a low-speed photodetector. Practically, this scheme is limited in terms of the number of channels and the receiving speed.

As is well known, OFCs can provide massive equally spaced wavelength channels, which can be employed as the multicarrier modulation, or combined with another similarly spaced comb to compose a dual OFC. Thus the development of optical frequency comb made a certain progress in channelized receiving schemes for broadband signals [17], [18], [19], [20], [21], [22], [23], [24], [25], [26], [27]. Recently, the novel OFCs based on microresonators have been produced by the Kerr effect in the high-Q resonators [28], [29], [30], [31], [32]. Extensive microcomb lines with the FSR ranging from tens of GHz to THz greatly increase the number of channels and operating bandwidth. It overcomes the problems of previous platforms, such as the low repetition frequency of mode-locked fiber combs and the limited number of electro-optic combs and promotes the development of channelization towards large bandwidth, low power consumption and integration.

In this paper, we present a comprehensive review of the recent advances in microcomb-based RF channelization. The paper is organized as follows. Section II introduces the basic principles of microwave photonic channelization and summarizes the related research into the following four categories: optical filter bank, multi-wavelength laser sources, dual optical frequency combs and frequency shifting/scanning. We devote recent advances in optical frequency combs and microcomb in terms of generation methods, nonlinear dynamics, fabrication processes, material platforms and applications in Section III. Focusing on the development of microcombs in microwave photonics, microcomb-assisted photonic channelization schemes are discussed. Section IV addresses the prospects for microwave photonic channelization. Photonic integrated circuits (PICs) for RF photonic channelizers are presented along with mature silicon-based photonic devices fabrication process and mainstream integration techniques, as well as current challenges and limitations. Finally, Section V summarizes the unprecedented advantages of microcomb-based RF channelization in terms of size, weight, and power (SWaP) reduction, and predicts that high-performance, chip-scale photonic channelized microsystems will revolutionize existing application scenarios.

## II. GENERAL APPROACHES

Microwave photonic channelization is designed to modulate RF signals into the optical domain, then use the abundant spectral resources in the optical domain and exploits optical approaches to split the broadband RF signal into multiple sub-narrowband signals for receiving and processing. As a result, this configuration overcomes the bandwidth limitations of ADCs/DACs, increasing the receiver's equivalent bandwidth and facilitating subsequent digital signal processing (DSP). In general, RF channelization schemes can be divided into four categories based on their main technique used to slice the broad spectrum: optical filter bank [9], [10], [33], [34], [35], [36], [37], [38], [39], multi-wavelength laser source [11], [12], [40], [41], [42], [43], [44], [45], [46], [47], [48], [49], [50], [51], [52], [53], dual optical frequency combs [13], [54], [55], [56], [57], [58], [59], [60], [61], [62], [63], [64], [65], [66], [67], [68] and frequency shifting/scanning [14], [15], [16], [69], [70], [71], [72].

The principles of the approaches are illustrated in Fig. 1; the experimental setups of representative technologies are shown in Fig. 2; and Fig. 3 shows a timeline of the evolution of RF channelizers, mainly in the past two decades. Detailed introductions of photonic RF channelization techniques are elaborated below.

### A. Optical Filter Bank

As shown in Fig.1 (a), the first scheme is to modulate the RF signal onto a single optical carrier and feed it into a spectrally dense optical filter bank. Thereafter, the broadband RF spectrum is divided into multiple subchannels in the optical domain and then down-converted to RF domain by photodetectors to obtain RF spectral slices, realizing multi-channel reception simultaneously. The implementation of the optical filter bank can be achieved via devices such as: acousto-optic crystals [9], diffraction gratings [41], fiber Bragg gratings [10], [34], Fabry-Perot filters [35], [39], arrayed waveguide grating [36], dual-polarization dual-parallel Mach-Zehnder modulator [37] or commercial waveshaper [38]. Initially, researchers proposed a channelization technique based on the surface acoustic wave, which diffracts the RF signal into waves with different spatial angles as they propagate through a phased array, resulting in 20 channels with 5 MHz channel spacing covering the 155 MHz-255 MHz band [9]. Later, the FP etalon was introduced to slice broadband RF signals, but the channel equalization was not ideal [39]. Phase-shifted chirped Bragg gratings, characterized by a transmission notch with different center wavelengths, were used to physically separate different RF spectral components [34]. Furthermore, the integration of a Bragg grating FP etalon and a hybrid Fresnel lens formed an optical channelization device to spatially separate the modulated microwave signals [10]. This solution provides 40 channels at 1 GHz bandwidth while compressing the system size.

This method features straightforward architectures and requires only one optical carrier, although it is not without challenges. First, the channelized receivers' performances (such as spectral slice resolution, bandwidth, adjacent channel leakage, and so on) are subject to the frequency response of the optical filters, thus which should feature narrow and flat passbands, steep roll-offs, and precise center wavelengths—challenging to implement in practice as the number of channels are relatively large. Second, due to the presence of insertion loss, the power attenuation of the optical signal after passing through a series of optical filters cannot be ignored.

### B. Multi-wavelength Laser Sources

As shown in Fig.1 (b), in the second photonic RF channelization scheme, the RF signal is first modulated by a set of equally spaced carriers (i.e., multi-wavelength laser sources) to obtain multiple replicas in the optical domain. Then narrowband optical filters with different center frequencies are used to split the signal into a series of subchannels.

So far, multi-wavelength laser sources can be provided by discrete laser arrays [46], [50], [73] and optical frequency combs [11], [12], [40], [44], [48], [49], [51], which produce multiple copies of the signal after broadcasting. In [73], four discrete lasers provide different wavelengths to form a four-channel receiver, and it is obvious that this approach can only achieve a limited number of channels, which is not conducive to high-capacity broadband communications. Later, OFCs were proposed, which can be realized by a variety of mechanisms, including fiber nonlinear effects [11], [40], time lens [12], [44], [47], [49], [51], [52] and mode-locked laser [42], [53]. For instance, multi-wavelength optical carriers have been achieved via the four-wave mixing (FWM) effect in highly nonlinear fibers [40]. Optical combs generated by cascaded electro-optic modulators are also widely used for RF channelization [12], [44]. OFCs are particularly attractive because it usually requires only one laser source and the comb spacing is tunable, making the receiving bandwidth flexible and reconfigurable. However, the above schemes for generating OFCs require many bulky discrete components, are difficult to integrate monolithically, and perform mediocrely in channel number and spectral resolution. At present, integrated microresonators are emerging as a more advanced platform that produces microcombs with larger free spectral ranges (FSRs) and complementary metal-oxide-semiconductor (CMOS) compatibility, making them an excellent choice for chip-scale photonic channelized receivers [74], [75].

In addition, spectral slicing can be performed by stimulated Brillouin gain spectrum [46], wavelength division multiplexers (WDMs) [49], optical delay lines [42], FP etalons [11], [12], [43], [45], [52], free-space diffraction elements [41] and microresonators [74], [75], which act similarly to the periodic filters. As Fig.1 (c) illustrates, the optical carriers (modulating the RF signal) and the LO combs are simultaneously incident at different angles onto the diffraction grating, which performs the spectral-to-spatial conversion, resulting in spectral separation and the formation of different channels. Each comb line is coupled with the corresponding sliced narrowband optical signal, which is then fed to the photodetector array for photoelectric conversion. We note that, for the approach based on optical delay lines, multiple tunable LOs can be generated from the beatnotes of the chirped pulses and their delays, which allows different frequencies to be down-converted without external signal sources; in addition, the number of channels can be improved by adding delay lines [42]. Later, a channelization scheme was proposed to introduce different fiber delays and attenuations to each subchannel, resulting in a designed magnitude response of each subchannel and enabling simultaneous spectrum slicing and sampling [53].

In summary, the multi-wavelength source method generally adopts coherent reception, which can preserve the phase information of the original RF signal, and is beneficial for the instantaneous spectrum analysis. Although we note that, it still requires various types of optical filters and faces the problems such as filter-to-filter frequency alignment and sensitivity to environment, resulting in low RF sensing accuracy and insufficient inter-channel suppression ratio.

### C. Dual Optical Frequency Combs

Dual optical frequency combs have been one of the most widely studied channelization schemes to date. As Fig.1 (d)

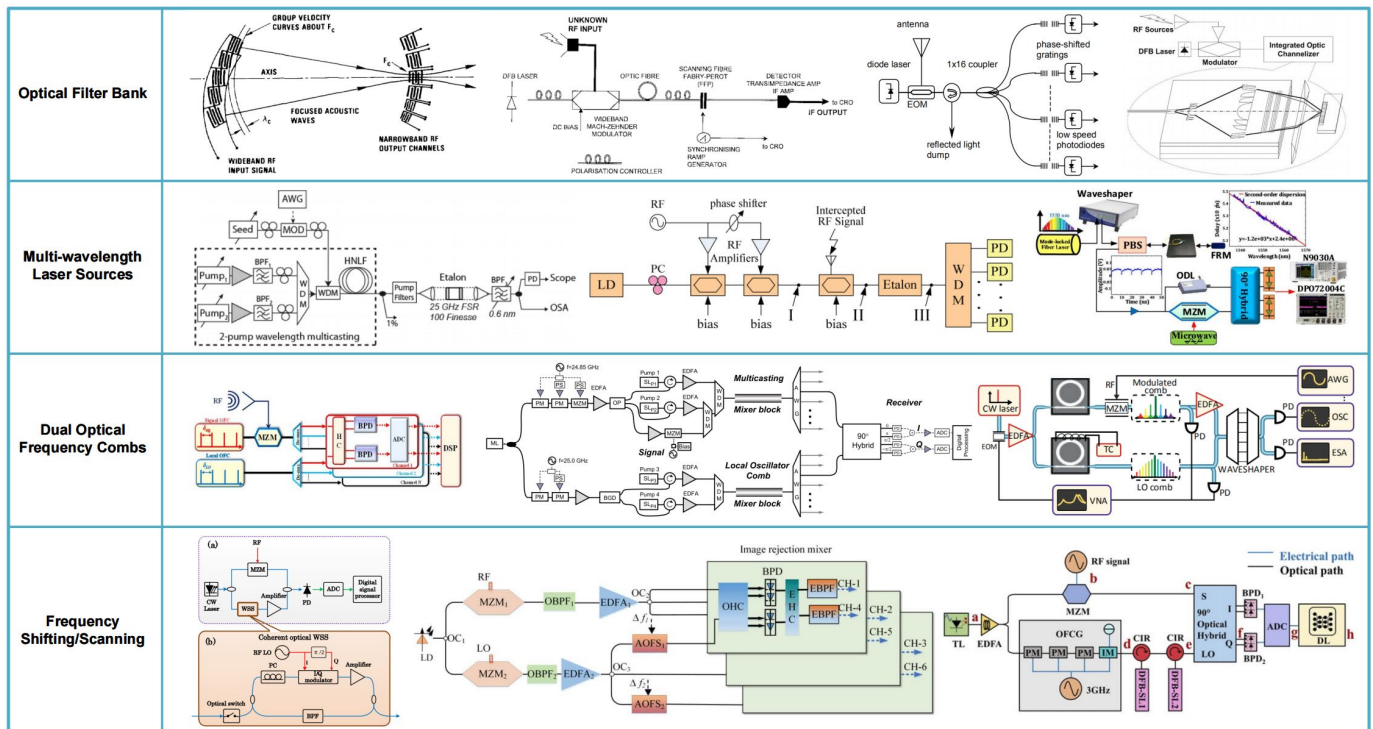

Fig. 2. Microwave photonic channelized receivers based on different techniques. Figures are adopted from [9], [35], [34], [10], [11], [12], [42], [13], [57], [60], [70], [15], [14].

depicts, two combs with slightly different FSRs (generally generated by cascaded electro-optic modulators [13], [59], [61], [62], [63], [64], [67] and microresonators [60]). One is used as carriers to multicast broadband RF signal, and the other serves as LOs, which is applied to probe the modulated RF components. The dual-comb vernier effect enhances the frequency sensitivity of the channelization and reduces the bandwidths requirements of the subsequent electronic devices. Each comb line corresponds to a channel, with the channel bandwidth equal to the frequency difference between the two combs. The two branches are mixed and fed into I/Q coherent detection, where the LO and the adjacent signal comb beat to obtain the corresponding RF components. Finally, the channelized spectrum is stitched together to recover the original broadband RF signal [47], [61].

In 2012, Xie et al. proposed a microwave photonic channelization technique based on a pair of electro-optic modulated OFC and coherent detection (i.e., I/Q demodulation) [13]. Since then, there have been numerous studies of similar structures. We know that the adoption of I/Q coherent detection at the receiving end can address the loss of RF signal phase information problem caused by the square-law detection of photodetectors (PDs). However, after I/Q coherent detection, the frequency components on the left and right of the LO are downconverted to the baseband (the desired signal is the useful signal, and the one symmetrical to the useful signal about the LO is image frequency) Typically, a Hartley structure image-reject mixer is introduced to achieve large instantaneous bandwidth and large in-band interference suppression [54], [59], [66], [71]. There are also some studies focusing on improving the performances of channelizer through spectrum stitching [61] and polarization multiplexing [66]. In addition, a dual optical comb can also be a pair of microcombs [60], or a microcomb and an electric comb [65], usually using the same continuous-wave (CW) pump to ensure coherence between them.

In conclusion, the dual-comb channelization scheme eliminates the need for optical filters, and the channel bandwidth (i.e., comb spacing) and channelized spectral resolution (FSR difference between the dual combs) can be flexibly adjusted. Importantly, coherent detection maintains a high signal-to-noise ratio (SNR). However, channel number is limited by the spectrum bandwidth of the optical comb, and massive comb lines require multiple electro-optic modulators, phase shifters, amplifiers, and microwave sources, which increases system complexity.

### D. Frequency Shifting/Scanning

In recent years, channelization schemes based on LO frequency shifting/scanning have been proposed, typically using acousto-optic frequency shifters (AOFS) [15] and injection locking [14], [69]. As shown in Fig. 1(e), a broadband RF signal is first modulated onto an optical carrier, while the LO comb is generated by electro-optic modulation. The external lasers are then injected into the LO comb to amplify the power of a single comb line. This wavelength is selected as a probe and fed into the coherent detection along with the modulated signal, and as the injection-locked LO comb line is changed, each portion of the modulated signal is sequentially down-converted to the intermediate frequency (IF) for subsequent digital processing.

In addition to the injection locking mentioned above, the LO frequency shifting can also be implemented by AOFS, where the shifted signal is fed to the I/Q coherent receiver along with the modulated signal, and finally the broadband RF signal is split into six subchannels for output [15]. In some studies, the round-trip time of the cyclic frequency-shifting loop was varied by controlling the duration of a single pulse through the optical switch to ensure that each frequency is within the appropriate time window, similar to wavelength scanning, to achieve multiple RF channels serial in the time domain [16]. Since there was an inherent frequency difference between the modulated RF signal and the frequency-shifted LO signal, the individual subchannels can be separated and down-converted to IF using I/Q coherent detection.

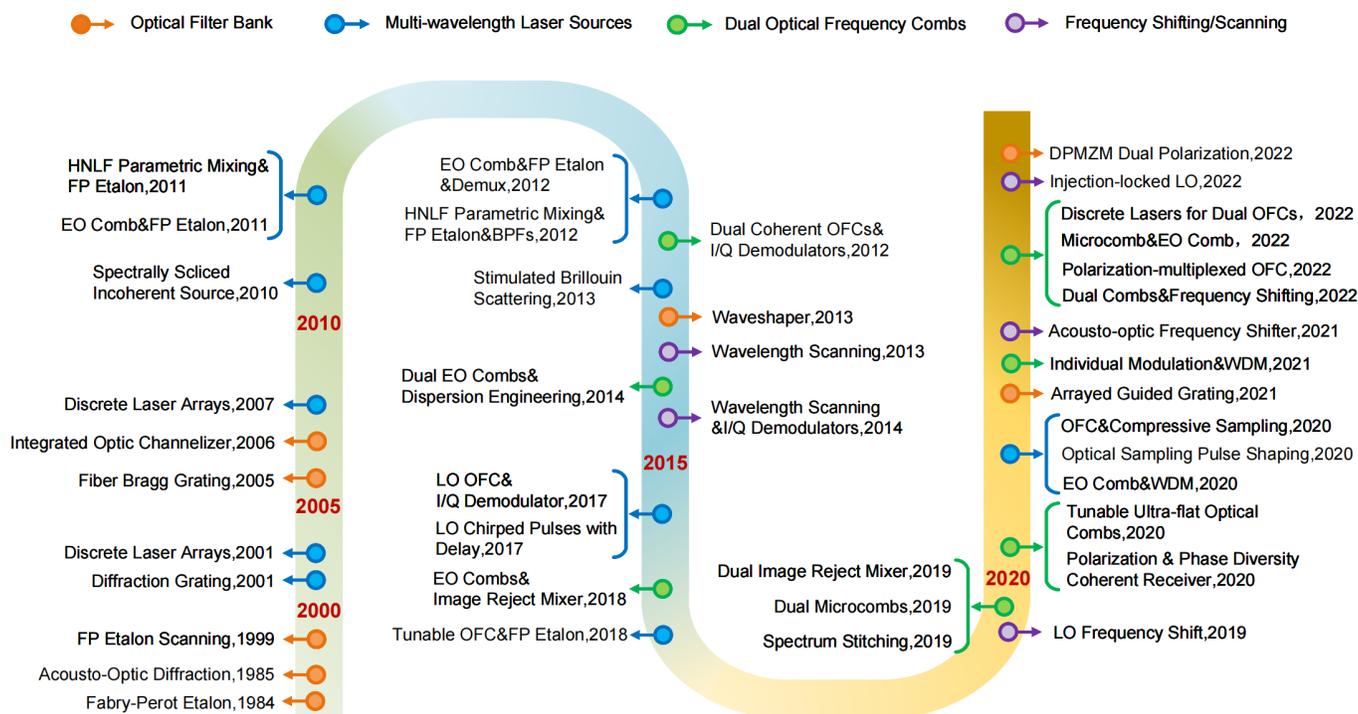

Fig. 3. Timeline showing the development of photonic-assisted channelization.

Single-carrier frequency-shifted channelization eliminates the need for ultra-narrow filters or coherent dual combs, simplifying the spectrum slicing process and allowing a relatively simple system architecture for real-time RF measurements. However, the number of channels is usually fixed and limited by the operating bandwidth of the frequency shifter or the wavelength scanning bandwidth.

More recently, researchers continue to focus on the optical combs, which are used to generate multiple optical carriers or local oscillators for RF signals broadcasting and frequency conversion, thus continuously improving the performance of microwave photonic channelization.

## III. MICROCOMB FOR PHOTONIC CHANNELIZATION

The optical frequency comb behaves as a sequence of equally spaced, phase-coherent modes in the frequency domain. Since it was first introduced, it has been regarded as the most effective tool for spectral measurement and has driven the rapid development of the spectroscopy and precision measurement. Meanwhile, offering a large number of equally-spaced wavelength channels within relatively compact footprints, OFCs are also promising light sources for microwave photonic channelization applications [76].

Mode-locked fiber combs are one of the most widely used OFCs [21], [27], [77], [78] in applications including precision metrology [79], frequency stabilization [80], optical frequency synthesis [81] and others. However, subject to the relatively long cavity (generally implemented by optical fibers), the spacing of comb lines generally on the order of MHz, which imposes limitations on the Nyquist bandwidth for RF spectra procession/channelization.

Another OFC scheme based on cascaded electro-optic modulators [18], [82] can offer flat comb lines with large spacings up to tens of GHz, and agile center wavelengths and repetition rates. However, the use of external microwave sources and high-power RF amplifiers (needed to obtain broader bandwidths) can significantly increase the cost and complexity of the system, unfavorable for mass production and generic applications.

Benefiting from advances in micro- and nanofabrication technologies, the researchers have innovated in the physical mechanisms of optical frequency comb generation as well as microresonator materials and process platforms. In 2003, a silica toroid-shaped microdisk with a quality (Q) factor of 1.25 $\times$ $10^8$ was fabricated for the first time through semiconductor photolithography and etching processes [83]. The following year, optical parametric oscillations ("microcomb prototype") were observed in the above microcavity [84]. Later, by further increasing the optical input power, a cascaded four-wave mixing effect was excited in the monolithic toroidal microresonator to form broadband equally spaced optical frequency combs [29]. Since then, the strictly microcombs was born. The dissipative Kerr soliton was soon proposed [85], which overcame the drawback of poor coherence of the previous optical combs. Octave-spanning microcombs covering the visible, near-infrared and mid-infrared bands have recently been obtained by dispersion engineering, combining $\chi^{(2)}$ and $\chi^{(3)}$ effects and dual pumping [86], [87], [88], [89], [90], [91], [92], [93], extending the bandwidth for *f-2f* self-reference technique and metrology.

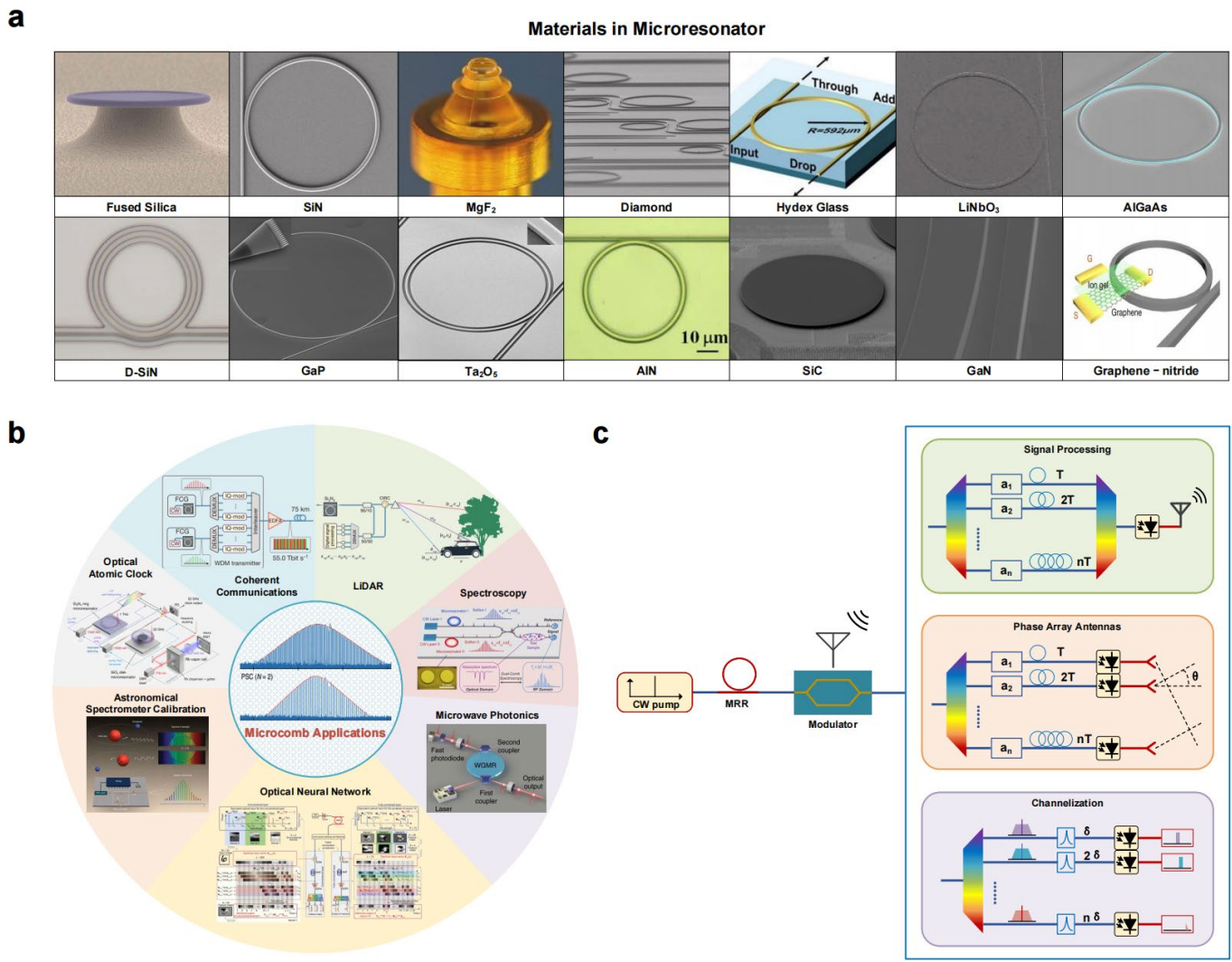

Fig. 4. (a) Microresonator material platforms, including fused silica [29], silicon nitride [112], MgF$_2$ [85] [113], diamond [114], Hydex glass [115], lithium niobate [116], [117], aluminum gallium arsenide [118] [119] [120], deuterated silicon nitride [121], gallium phosphide [122], tantalum pentoxide [123], aluminium nitride [124], [125], silicon carbide [126], gallium nitride [127], graphene nitride [128]. (b) Applications based on microcombs, including coherent communications [137], light detection and ranging (LiDAR) [138], spectroscopy [139], [140], optical atomic clock [141], astromical spectrometer calibration [142], optical neural network [143], [144] and microwave photonics [145], [146], [147], [148], [149]. (c) Microwave photonic classical applications based on microcombs, including signal processing [150], [151], [152], [153], [154], [155], phase array antennas [156], [157] and channelization [74], [75]. Figures are adapted from [29], [112], [85], [114], [115], [116], [118], [121], [122], [123], [124], [126], [127], [128], [137], [138], [139], [141], [142], [143], [146].

For photonic RF channelization, microcombs bring many unique advantages, including high repetition rates, small footprints, low power consumption and CMOS-compatible fabrication processes for mass production. Specifically, photonic RF channelizers' operating bandwidth (i.e., the Nyquist zone) is equal to half of the comb spacing and represents the maximum microwave frequency that can be processed without interference from adjacent wavelength channels. Therefore, the large FSRs of microcombs (ranging from tens of GHz to THz) enable channelization/processing of broadband RF signals [94], [112].

According to the Lugiato-Lefever equation (LLE), microcomb generation results from a double balance between dispersion and nonlinearity as well as loss and parametric gain [95]. When the pump sweeps rapidly from the blue detuning of a resonance to the red side, the cascaded FWM effect occurs in the resonator, accompanied by an evolution from the primary comb, secondary comb, modulation instability comb to soliton comb [85]. The matching of pump power and frequency detuning is critical for the formation of stable soliton combs. Later, many soliton states have been observed, such as bright soliton [85], soliton crystal [96], soliton molecule [97], breathing soliton [98], [99], dark soliton [100], Stokes soliton [101], soliton Cherenkov radiation [102], and so on. Moreover, various physical phenomena have also been discovered, such as avoided mode crossing [103], Raman self-frequency shift [104], and second harmonic generation [105], [106], [107], which greatly enrich the nonlinear dynamics of microresonators.

Low-noise coherent microcombs exhibit essential for practical applications. Whereas, the main problem is the limitation of the thermo-optic effect. Currently, many schemes have been demonstrated to essentially stabilize the pump light at the red detuning of the resonance, such as fast frequency tuning [85], "power kicking" [108], [109], thermal tuning [110], auxiliary laser heating [111], etc. The above approaches are suitable for different application scenarios. As we know, the pump scanning to the red detuning of the resonance will reach the single-soliton state, at which point the intracavity power decreases. To obtain a steady single-soliton microcomb, one idea is to introduce auxiliary light on the blue side of the resonance to increase the auxiliary light power, thereby maintaining the thermal balance of the microcavity.

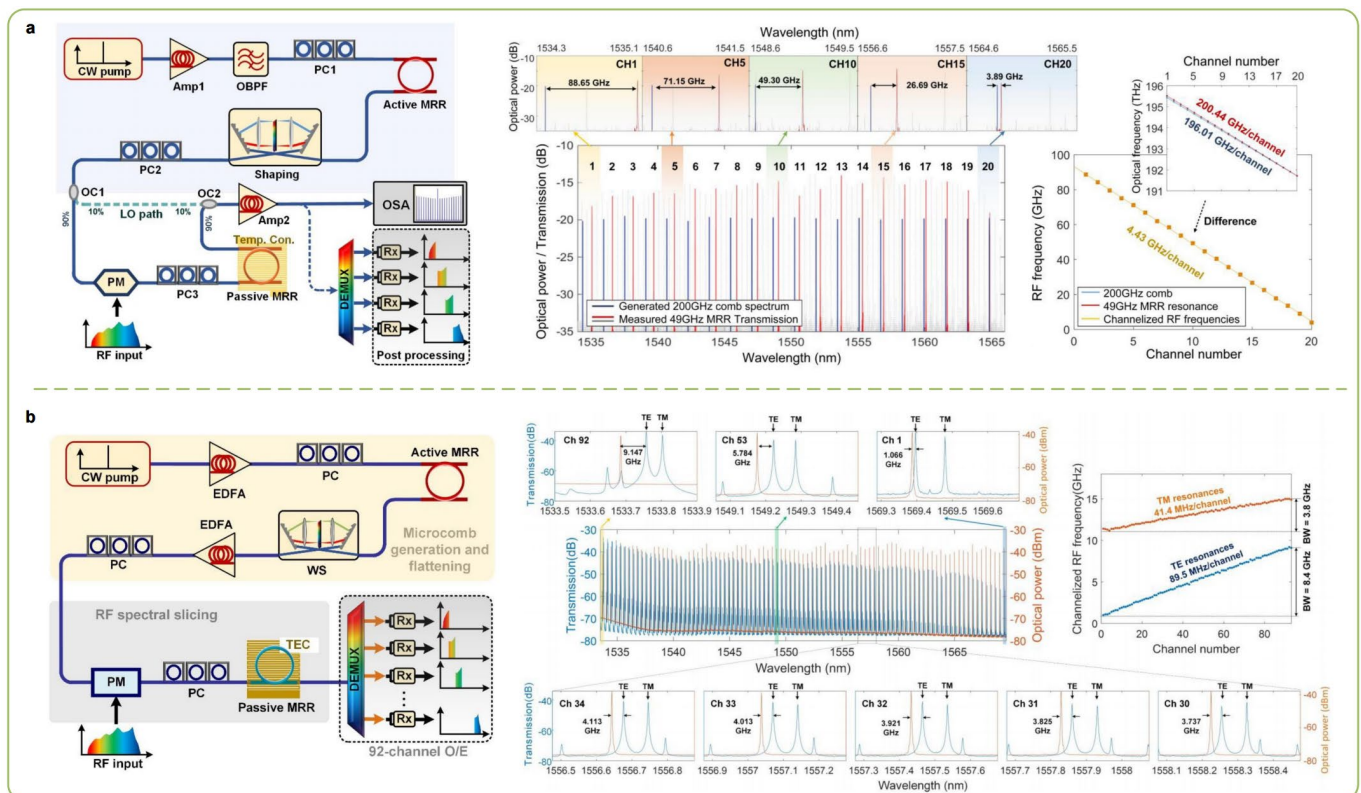

Fig. 5. Previous work based on microcomb [74], [75].

Microresonators are mainly fabricated from fused silica [29], SiN [112], MgF$_2$ [85], [113], diamond [114], Hydex glass [115], LiNbO$_3$ [116], [117], AlGaAs [118], [119], [120], D-SiN [121], GaP [122], Ta$_2$O$_5$ [123], AlN [124], [125], SiC [126], GaN [127] and graphene nitride [128], as Fig. 4 (a) shows. They are manifested in various forms such as microdisks [129], microrods [130], microspheres [131], microtoroids [83] and microrings, etc. As for crystalline resonators，such as CaF$_2$ [132], [133], MgF$_2$ [134] and fused silica exhibit ultra-high Q factor of $10^8$-$10^{10}$. For monolithic integrated microcombs, Si$_3$N$_4$, Hydex glass, LiNbO$_3$ and other materials are widely applied on chip, limited by waveguide transmission losses, Q values range from $10^5$ to $10^7$. These features have high Kerr nonlinearity and low propagation loss, and are compatible with CMOS platforms. However, microcomb generation requires an external laser source and discrete optical and electrical components which does not significantly reduce system complexity. Heterogeneous integration is expected to solve the above longstanding problems [135], and studies have been conducted to heterogeneously integrate indium phospide/silicon (InP/Si) semiconductor lasers and SiN microcombs on monolithic silicon substrates to realize laser soliton microcombs. Subsequently, a parallel optical data link and a highly reconfigurable microwave photonic filter were also demonstrated on the aluminium-gallium-arsenide-on-insulator (AlGaAsOI) platform with a high third-order nonlinear coefficient ($n_2 \approx 2.6 \times 10^{-17}$ m$^2$W$^{-1}$) and a Q factor >2 million [136]. Therefore, the turnkey dark-pulse comb requires a rather low threshold at the few-milliwatts level, which can be provided by a DFB laser chip. The state-of-the-art heterogeneously integrated III-V lasers and microresonators have inspired researchers to move more applications to on-chip, promising full integration of optical systems in the future.

As shown in Fig.4 (b), dissipative Kerr microcombs have already enabled advances in massive coherent optical communications [137], light detection and ranging (LiDAR) [138], spectroscopy [139], [140], atomic clocks [141], astronomical spectrographs calibration [142], optical neural network [143], [144] and microwave photonics [145], [146], [147], [148], [149]. Here, we discuss the microwave photonics applications based on integrated microcombs, as Fig.4 (c) illustrates. A microwave photonic system generally consists of a laser source, a modulator, a signal processing unit, and a photodetector. Specifically, the microcomb serves as a multi-wavelength laser source, which promotes the development of integrated microwave photonics. Research in microwave photonics has focused on signal processing (including microwave signal generation [150], [151], photonic RF filters [152], [153], intensity differentiator [154], Hilbert transformers [155] etc. ), phased-array antennas [156], [157] and channelizers [74], [75]. The signal processing typically involves flattening the generated microcombs, feeding them into a modulator to broadcast microwave signals, and then introducing different time delays for each comb by a spool of dispersive fiber. The optical comb lines are reshaped by assigning the designed weights, which after photodetection can produce arbitrary transfer functions such as photonic filters, Hilbert transformers and differentiators. Compared to electro-optic frequency combs or mode-locked fiber combs, microcombs feature a much larger mode spacing (hundreds of GHz), corresponding to a larger Nyquist operating bandwidth (half the FSR of a microresonator). In addition, coherent combs yield low phase noise photonic microwave oscillators [158], [159] and frequency synthesizers [160], [161], [162]. For the phased-array antennas, the microcombs are used as true time delay lines (81 channels in the C-band) for modulating microwave signal copies at each wavelength, and then dispersion is introduced by the single-mode fiber (SMF) to generate the time-delay difference between adjacent channels [157]. Finally, a programmable optical filter and a wavelength division multiplexer (WDM) are used to separate each channel, and the microwave signals with different delays are fed to the antenna array after photodetection, thus realizing the arbitrary beam patterns. Another typical application of the microcombs is RF photonic channelization, where a broadband microwave signal is modulated on both sides of the comb lines, then sliced through a narrowband filter bank such as a high-Q microresonator (the FSR is slightly different from the active one), and finally sent to the photodetection. Dual-coherent microcombs can also perform a similar effect to obtain the amplitude and phase information of the microwave signal.

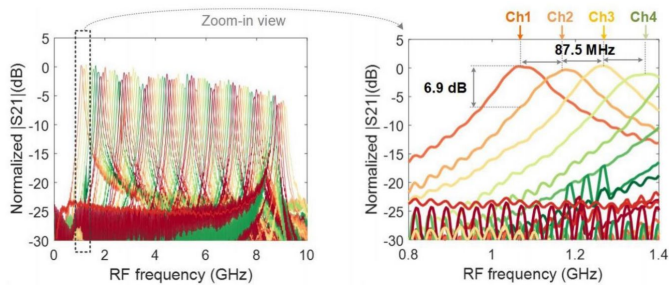

Fig. 6. Previous work based on microcomb [75].

In previous work [74], we used an active nonlinear MRR to generate broadband comb spectra with a 200 GHz spacing, and then flattened the 20 Kerr comb channels on the C-band (limited by the bandwidth of the waveshaper in this experiment, there are actually more than 60 channels on the C+L band) with a programmable optical filter. As shown in Fig. 5(a), we introduced a spectrometer-waveshaper feedback control path, in which the microcomb power is first detected by an optical spectrum analyzer and then compared to an ideal (i.e., equal) value, generating an error signal that is fed back into the waveshaper for cyclic comb shaping. The input RF signal is then multicast onto each comb using a phase modulator. Next, the RF signals on each wavelength channel are spectrally sliced by every fourth resonance of a passive MRR with an FSR of 49 GHz and a Q factor of $1.55 \times 10^6$. The RF channelization resolution is 1.04 GHz, compatible with state-of-the-art ADCs. Later, the 20 channels and the transmission spectrum of the passive MRR was measured, noting that the relative offset between the optical comb and every fourth resonance varies from 3.89 to 88.65 GHz, so that the channelized RF bandwidth was close to 90 GHz. The two are linearly fitted to obtain the channelized RF frequency step of 4.43 GHz per channel. We experimentally demonstrated the RF performance at four wavelengths in 20 channels, corresponding to four RF frequencies from 1.7 to 19.0 GHz—close to the limit of our device (up to 20 GHz). However, with only four channels operating simultaneously, the channelized RF step is too large, resulting in discontinuities in the received RF signal and limiting the overall instantaneous bandwidth (the product of the number of channels and the slicing resolution). Finally, these "gaps" are compensated by thermally tuning the passive MRR.

Later, we further improved this scheme [75]. As shown in Fig. 5(b), using two MRRs with nearly matched FSRs (~ 49 GHz), and the active MRR generates a soliton crystal comb providing up to 92 channels (C-band), and the passive one is used as a periodic band-notch filter. Utilizing the Vernier effect between the two, the broadband RF spectrum was sliced at a high resolution of 121.4 MHz to achieve continuous channelization with an RF channelization step of 87.5 MHz/channel. Finally, we experimentally verified an instantaneous RF operating bandwidth of 8.08 GHz—more than 22 times higher than the previously reported bandwidth, and achieved RF channelization up to 17.55 GHz with the aid of thermal tuning. Similarly, we measured the 92 optical combs and passive MRRs transmission spectra (containing both TE and TM polarization resonances), and extracted that the total operating bandwidths of the TE and TM resonances are 8.4 GHz and 3.8 GHz, respectively. And the RF channelization steps for each channel are 89.5 MHz (TE) and 41.4 MHz (TM), respectively. After passive MRR, the RF spectrum is divided into multiple segments transmitted on different channels, which are demultiplexed and detected separately and the results of 92 parallel RF channels are measured by the vector network analyzer, as shown in Fig. 6. This microcomb-based approach is a significant step toward the full integration of photonic receivers into modern RF systems, offering the advantages of massively parallel channels, high resolution and well suited for broadband signal detection and processing.

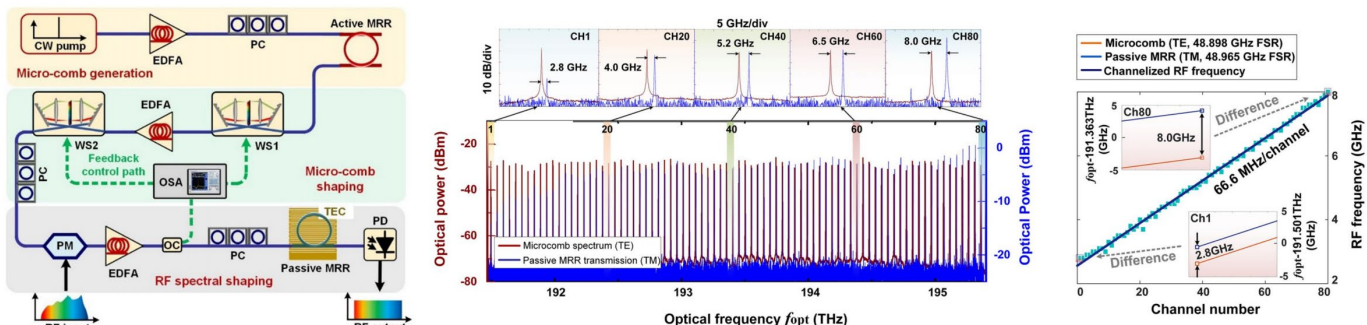

Fig. 7. Previous work based on microcomb [163].

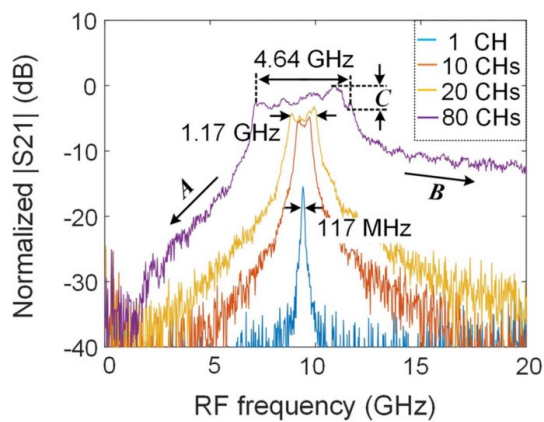

Fig. 8.  Previous work based on microcomb [163].

Based on the scheme [75], a programmable arbitrary transfer function RF filter was implemented by adding a second-stage microcomb shaping and using a single PD to combine channelized RF components during photodetection [163], as shown in Fig. 7. We measured the flattened microcomb of 80 channels and the transmission spectrum of the passive MRR, and fitted to obtain a channelized RF frequency step at 66.6 MHz/channel, covering the sliced RF frequencies from 2.8 to 8.0 GHz. As shown in Fig. 8, the RF filter operating bandwidth extends from 1.17 GHz to 4.64 GHz with a resolution of 117 MHz as the channel number increases from 20 to 80. The high-resolution photonic RF filters have been demonstrated by utilizing an RF bandwidth scaling method based on the integrated Kerr microcomb.

## IV. OUTLOOKS OF MONOLITHIC INTEGRATION

High-bandwidth, multi-channel, high-resolution, fully on-chip integration and turnkey operation are becoming the trend for next-generation microwave photonic channelized receivers [147], [164], [165], [166], [167], [168]. In the current advanced demonstration of microcomb channelization, only the micro-ring resonator is chip-scale as a stand-alone integrated device, while the other components are discrete devices [74], [75]. Therefore, it is important not only to develop the integration of individual device units, but also to facilitate their integration on the same platform to realize photonic integrated circuits (PICs) [169], [170], [171]. Over the years, many PICs based on various waveguide materials such as silicon-on-insulator (SOI) [172], silicon nitride (SiN) [173], indium phosphide (InP) [174], lithium niobate (LiNbO$_3$), aluminum nitride (AlN) [175] and silicon carbide (SiC) [176] have been investigated in academia and industry. Among these, silicon is the most widely used waveguide material. Silicon-based photonics has the benefit of CMOS process compatibility and the potential to enable monolithic integration of silicon-based electronics and photonics with high quality, low cost, and high throughput with the assistance of the mature manufacturing processes. The fact that silicon has been considered an "electrical" rather than an "optical" material [177]. This is its own limitation, that is, as an indirect energy bandgap semiconductor material, silicon requires the participation of phonons to fulfill the momentum conservation, making it a very inefficient light source. Not only that, the silicon lattice has central symmetry, and there is almost no electro-optical effect. Thus, it seems that silicon material is not a good choice either as a light source or an electro-optical modulator or photodetector [178]. Other waveguide materials can overcome some of the drawbacks of silicon, but they are potentially limited in other ways, and the dominant solution is silicon-based multi-material fusion. This allows each material to perform optimally for its photonic component without compromising the functionality of other components in the system. The goal remains to realize silicon photonic devices on a chip, co-packaged with integrated lasers that interface directly with the silicon chip for a variety of advanced applications [136]. Over time, three integration processes have emerged: hybrid integration, heterogeneous integration, and fully monolithic integration [179], [180]. These are briefly described below.

### A. Hybrid Integration

Hybrid integration is an integration process that typically combines two or more PIC or photonic device chips with different functions from different material technologies in a single package [165], [170], [179], [180], [181]. The advantage is the ability to test and characterize the devices to be integrated before the integration, and to easily select the components to be integrated into photonic circuits on a case-by-case basis without having to consider the compatibility of device materials. In particular, the hybrid integration of fully processed III-V devices on silicon and silicon nitride PICs has solved the problem of silicon not emitting light, enabling structures such as lasers, modulators and even photodetectors [164], [182], [183], [184], [185], [186], [187]. However, most hybrid integrated III-V/SiN photonic devices still require one-to-one alignment (i.e., light waves from different substrates must be aligned at their respective interfaces to improve the efficiency of inter-chip optical coupling), which limits large-scale production and packaging, resulting in with lower efficiency and improved reliability.

### B. Heterogeneous Integration

Heterogeneous integration is an integration process that refers to the technology of combining different materials (devices) on a single substrate through wafer bonding, typically in the early- to mid-stages of PIC chip fabrication [170], [179], [180], [181]. The heterogeneous integration of silicon-based and III-V devices has been a hot topic over the past few years [136], [166], [178], [188], [189]. Unpatterned III-V thin films are integrated onto pre-processed silicon photonic wafers and the devices are then lithographed over the entire wafer scale. Due to the mismatch between the effective modal indices of active gain materials (e.g., InP, GaAs, etc.) and passive waveguides (ultra-low loss SiN), Si is typically used as an interlayer to bridge the modal refractive indices. The seamless integration of the two through the multimodal hopping mechanism takes advantage of their respective strengths to realize a versatile heterogeneous integrated III-V/SiN platform that does not need to rely on an external light source. Heterogeneous integration moves the manufacturing process from the chip level to the auto wafer level, provides high alignment accuracy and low loss when switching between different waveguide materials, and improves the reliability and integration density, making it suitable for high-volume production and applications. However, heterogeneous integration does not allow performance testing of each component in advance, placing more stringent requirements on the wafer bonding process to improve throughput.

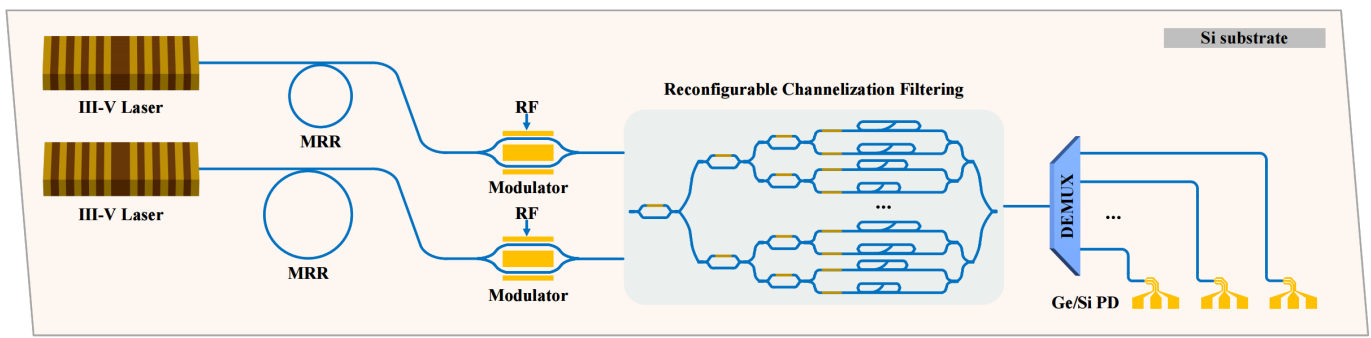

Fig. 9. Prospect view of heterogeneous integrated microwave photonic channelized receiver.

### C. Fully Monolithic Integration

Fully monolithic integration is the solution to the lack of a core light source for silicon-based integrated circuits. It involves the integration of III-V gain materials on a silicon substrate by direct epitaxial growth [170], [179], [181]. In addition, the entire process does not involve the bonding of materials, which is suitable for large-scale growth and high-volume production, reduces the cost of the integrated system, and provides a very broad application prospect. However, due to the polarities, lattice constants and thermal expansion coefficients mismatch between the III-V materials and silicon materials, direct growth of III-V materials on silicon can lead to threading dislocations, stacking faults, misfit dislocations and antiphase domains (APDs), which seriously affects the operating performance and lifetime of devices [181], [190]. To address this problem, the researchers used a quantum dot structure that is insensitive to APDs and dislocation defects to produce a long-lived, high-performance laser (operating near 1.3 um), which is expected to be the core light source for photonic circuits [185], [191], [192], [193].

However, we must recognize that fully monolithically integrated silicon-based photonic circuits and devices are still in their infancy, and there is much to be done for future research. Heterogeneous integration, on the other hand, is a relatively mature technology that can diversify the functionality of a single chip while maintaining the original device and process dimensions. Lately, some exciting work has been carried out by researchers using multilayer monolithic integration and heterogeneous integration in concert [168]. It can be said that silicon-based monolithic heterogeneous integration will open a new way for the development of microelectronics, optoelectronics and microsystem technologies in the post-Moore era.

### D. Integration of Photonic RF channelizers

Here, we propose several perspectives for the future of channelized receivers based on the heterogeneous integration scheme, as shown in Fig. 9. The microcomb generation draws on the scheme and processing technology in [135], which integrates an InP/Si continuous-wave laser and a high-Q SiN MRR on a monolithic silicon substrate. Utilization of back-Rayleigh scattering from the microresonator to the laser for self-injection locking and control of the current and phase to achieve optimal pump-resonance detuning results in the formation of a Kerr soliton microcomb (the thermo-optic phase tuner and electronic control circuitry are omitted from the figure). In recent years, electro-optic modulators based on lithium niobate films [194], [195], graphene [196], [197], silicon [198], [199], [200] and other materials have emerged. Given the maturity of silicon photonics processes, silicon modulators are used here, which typically rely on the dispersion effect of free carrier plasma, and electro-optic bandwidths above 50 GHz have been investigated [199]. Moreover, mature silicon-based Mach-Zender modulators (MZMs) or micro-ring modulators (MRMs) can be integrated with SiN photonic circuits in multiple layers [186], [201], and the ultra-low transition loss of the Si-SiN layer helps silicon modulators achieve high-speed and effective modulation. Channelized filtering is realized using an MZI waveguide mesh composed of integrated silicon-based tuning elements, and a reconfigurable FIR filter is implemented in combination with hardware programming [202], [203]. The optical signal passes through the filter unit and then enters the demultiplexer to be divided into independent channels, each of which is coupled to an on-chip germanium photodetector to achieve signal reception [204]. Such a silicon photonic integrated circuit is compatible with mature CMOS fabrication technology, which can take advantage of the superior performance of different materials and realize more new ideas to develop turnkey, high-bandwidth and scalable integrated microwave photonic channelizers.

## V. CONCLUSION

In this paper, we review the current state of the art in microwave photonic channelizers and discuss the advantages and shortcomings of the four technology paths, namely optical filter banks, multi-wavelength light sources, dual optical combs and frequency shifting/scanning. The idea is then presented that RF photonic channelization will move toward miniaturization and integration in the future. Starting with the key component, the microresonator, we introduce various microwave photonic applications based on microcombs. The outstanding benefits and potential applications of microcombs for RF photonic channelization are demonstrated in conjunction with previous studies. Finally, we outline the prospects for a silicon-based heterogeneous integrated RF channelized receiver using photonic integrated circuits and silicon photonics, aiming to create a high-bandwidth, low-cost and power-efficient solution to make it an attractive platform.

**Weiwei Han** is currently working toward the Ph.D. degree with the State Key Laboratory of Information Photonics and Optical Communications, Beijing University of Posts and Telecommunications (BUPT), Beijing, China. Her research interests include microcombs applications, and microwave photonics signal processing.

**Zhihui Liu** is currently working toward the Ph.D. degree with the State Key Laboratory of Information Photonics and Optical Communications, Beijing University of Posts and Telecommunications (BUPT), Beijing, China. His research interests include integrated Kerr microcombs and microcombs applications.

**Yifu Xu** is currently working toward the Ph.D. degree with the State Key Laboratory of Information Photonics and Optical Communication, Beijing University of Posts and Telecommunications. His research interests include integrated Kerr microcombs and optical computing.

**Mengxi Tan** (Student Member, IEEE) received the Ph.D. degree from the Swinburne University of Technology, Melbourne, VIC, Australia, in 2021. She is current an Associate Professor with Beihang University. Her research interests include microcombs and microwave photonics.

**Chaoran Huang** received the Ph.D. degree from the Chinese University of Hong Kong, China in 2016. She is currently an Assistant Professor with the Chinese University of Hong Kong, China. She was the recipient of 2019 Rising Stars Women in Engineering Asia, and was nominated by Princeton University to compete for Blavatnik Regional Awards for Young Scientists. She has authored more than 40 peer-reviewed research papers, three book chapters, and one US patent. She has been a TPC member of several international conferences and is a frequent reviewer for different journals in IEEE, OSA, and the Nature Publishing Group.

**Jiayang Wu** (Senior Member, IEEE) the Ph.D. degree in electronics engineering from Shanghai Jiao Tong University, Shanghai, China, in December 2015. He is currently a Senior Research Fellow and a Senior Lecturer with the Optical Sciences Centre, Swinburne University of Technology. His research interests include integrated photonics, nonlinear optics, 2D materials, and optical communications.

**Kun Xu** received the B.Sc. degree in applied physics from the Central South University of Technology (currently Central South University), China, in 1996, the M.Sc. degree in optical engineering from the University of Electronic Science and Technology, China, in 1999, and the Ph.D. degree in physical electronics from Tsinghua University, China, in 2002. Then he joined the Beijing University of Posts and Telecommunications, Beijing, China. He was a Visiting Scholar at Nanyang Technological University, Singapore, in 2004. He is currently the President of Beijing University of Posts and Telecommunications. His research interests include fiber-wireless integrated networks, distributed antenna systems, ubiquitous wireless sensing and access, RF photonic integrated systems, artificial intelligence and photonics neural networks, machine learning, and ultrafast optics. Prof. Xu has served for the Technical Program Committees (TPC) and the Workshop/Session Co-Chair of several international conferences, including the IEEE Microwave Photonics/Asia Pacific Microwave Photonics Conference (MWP/APMP), the IEEE Global Symposium on Millimeter Waves (GSMM), the IEEE International Conference on Communications (ICC), and Progress in Electromagnetics Research Symposium (PIERS). He was also the General Chair of 2020 IEEE/OSA/SPIE Asia Communications and Photonics Conference (ACP 2020), the TPC Co-Chair of ACP 2015 and ACP 2018, and the LOC Chair of ACP 2013 and APMP 2009. He was a Guest Editor of the Special Issue on Microwave Photonics in Photonics Research (OSA/CLP) in 2014.

**David J.Moss** (Life Fellow, IEEE) received the B.Sc. degree from the University of Waterloo, Waterloo, ON, Canada, and the Ph.D. degree from the University of Toronto, Toronto, ON, Canada. He is currently the Director of the Optical Sciences Centre, Swinburne University of Technology, Melbourne, VIC, Australia, and the Deputy Director of the ARC Centre of Excellence for optical microcombs for breakthrough science. He was with RMIT University, Melbourne, from 2014 to 2016, The University of Sydney, Camperdown, NSW, Australia, from 2004 to 2014, and JDSUniphase, Ottawa, ON, Canada, from 1998 to 2003. From 1994 to 1998, he was with the Optical Fiber Technology Centre, Sydney University, from 1992 to 1994 with Hitachi Central Research Laboratories, Tokyo, Japan, and from 1988 to 1992 with the National Research Council of Canada in Ottawa. His research interests include optical microcombs, integrated nonlinear optics, quantum optics, microwave photonics, ONNs, optical networks and transmission, 2D materials for nonlinear optics, optical signal processing, nanophotonics, and biomedical photonics for cancer diagnosis and therapy. He was the recipient of the 2011 Australian Museum Eureka Science Prize and Google Australia Prize for Innovation in Computer Science. He is a Fellow of the IEEE Photonics Society, Optica (formerly the OSA), and the SPIE.

**Xingyuan Xu**
Xingyuan Xu received the Ph.D. degree from the Swinburne University of Technology. He is currently a Professor with the Beijing University of Posts and Telecommunications, Beijing, China. His research interests include neuromorphic optics, optical signal processing, and optical frequency combs.